\documentclass[letterpaper,10pt]{article}
\usepackage{opex3}

\usepackage[T1]{fontenc}
\usepackage[ngerman, english]{babel}	
\usepackage{amsmath, amssymb}
\usepackage{floatrow}
\usepackage{cite}

\usepackage[labelsep=period]{caption}

\usepackage{subfig}

\usepackage{amsthm}
\usepackage{thmtools} 

\declaretheoremstyle[%
  spaceabove=0pt,%
  spacebelow=0pt,%
  headfont=\normalfont\itshape,%
  postheadspace=0.5em,%
  qed=\qedsymbol%
]{mystyle} 
\declaretheorem[name={Proof},style=mystyle,unnumbered]{MyProof}

\newtheoremstyle{MyPlain}%
{0pt}
{0pt}
{\itshape}
{1em }
{\bfseries}
{:}
{.5em}
{ }
\newtheoremstyle{MyDefinition}%
{0pt}
{0pt}
{}
{1em }
{\bfseries}
{:}
{.5em}
{ }
\newtheoremstyle{MyRemark}%
{0pt}
{0pt}
{}
{1em }
{\itshape}
{:}
{.5em}
{ }

\theoremstyle{MyPlain}
\newtheorem{lemma}{Lemma}

\theoremstyle{MyDefinition}

\theoremstyle{MyRemark}

\usepackage{xspace} 

\usepackage[
	ruled,
	vlined,
	boxed, 
	linesnumbered, 
	commentsnumbered]{algorithm2e}
\DontPrintSemicolon

\usepackage[
	shortcuts,
	nonumberlist,	
	acronym, 
	hyperfirst=true,
	section=section, 
	order=letter,
	numberedsection=nolabel 
]{glossaries}

\newglossary[slg]{los}{syi}{syg}{List of Symbols} 
\makeglossaries 
\glsdisablehyper 
\newglossarystyle{mylist}{
	\glossarystyle{list} 
}

\newglossarystyle{mysymbolstyle}{

	\renewcommand*{\glsgroupheading}[1]{}%
}

\newglossarystyle{myacronymstyle}{
  \glossarystyle{mysymbolstyle} %

}
\newcommand{\NewS}[5][\newcommand]{
	\newglossaryentry{symb:#2}{
		name=\ensuremath{#3},
		description={\nopostdesc #4}, 
		sort=sym#5,
		type=los,
	}
	\glsadd{symb:#2}
	\expandafter\def\csname #2\endcsname{\ensuremath{#3}} 
}

\newcommand{\NewC}[5][\glsadd]{
	\newglossaryentry{symb:#2}{
		name=\ensuremath{#3},
		description={\nopostdesc #4}, 
		sort=sym#5,
		type=los,
	}
	\expandafter\def\csname #2\endcsname{\ensuremath{#3}} 
}

\newcommand{\NewF}[6][\newcommand]{
	\newglossaryentry{symb:#2}{
		name=\ensuremath{#3\cdot#4},
		description={\nopostdesc #5}, 
		sort=sym#6,
		type=los,
	}
	\expandafter\def\csname #2\endcsname ##1{\ensuremath{#3{##1}#4}} 
}
\NewS{TXR}{R_0}{transmitted radius}{rx}
\NewS{txR}{r_0}{transmitted radius}{rx}

\NewS{TXP}{\Theta_0}{transmitted phase}{tx}
\NewS{txP}{\theta_0}{transmitted phase}{tx}

\NewS{RXR}{R}{received radius}{ry}
\NewS{rxR}{r}{received radius}{ry}

\NewS{RXP}{\Theta}{received phase}{ty}
\NewS{rxP}{\theta}{received phase}{ty}

\NewS{TXRdet}{\hat{R}_0}{transmitted radius}{rx}

\NewS{estX}{\hat{X}}{estimated symbol}{xh}
\NewS{esttxR}{\hat{R}_x}{estimated transmitted radius}{rxe}
\NewS{esttxP}{\hat{\Theta}_x}{estimated transmitted phase}{txe}

\NewS{NLPN}{\Phi_{\text{NL}}}{nonlinear phase noise}{phi}
\NewS{Len}{{L}}{total fiber length}{L}
\NewS{Power}{{P}}{input power}{L}

\NewS{X}{X}{transmitted symbol}{x}
\NewS{Y}{Y}{received symbol}{y}
\NewS{setX}{\mathcal{X}}{signal constellation}{xs}
\NewS{setAPSK}{\setX_{\text{APSK}}}{APSK signal set}{xsapsk}

\NewS{rad}{r}{radius}{r}
\NewS{ppr}{l}{points per ring}{l}
\NewS{phaseoffset}{\phi}{phase offset per ring}{p}
\NewS{NumRings}{N}{number of rings in the APSK constelllation}{N}
\NewS{RD}{\bold{r}}{radii distribution}{rr}

\NewS{NumSpans}{K}{number of spans}{k}

\NewS{SEP}{\text{SEP}}{symbol error probability}{ty}

\NewS{labeling}{\mathbb{L}}{labeling}{l}
\NewS{BRGC}{\mathbb{G}}{binary reflected gray code}{gc}
\NewS{ODP}{\otimes}{ordered direct product}{odp}

\NewS{Reg}{\mathcal{R}}{compact bounding region}{reg}

\NewS{imag}{\jmath}{imaginary unit}{j} 

\NewC{Code}{\mathcal{C}}{code}{C:code}
\NewC{vecI}{\boldsymbol{I}}{identity matrix}{i}
\NewC{vecZero}{\boldsymbol{0}}{all-zero vector}{0z}
\NewC{xor}{\oplus}{exclusive or}{0excl}
\NewC{SNR}{\text{SNR}}{signal to noise ratio}{s}
\NewC{mod}{\;\operatorname{mod}\,}{modulo operation}{m}	
\NewC{cconv}{\circledast}{circular convolution}{0}	

\NewC{natural}{\mathbb{N}}{set of natural numbers}{n} 
\NewC{real}{\mathbb{R}}{set of real numbers}{rs}
\NewC{rational}{\mathbb{Q}}{set of rational numbers}{q}
\NewC{integer}{\mathbb{Z}}{set of integer numbers}{zzzz}
\NewC{complex}{\mathbb{C}}{set of complex numbers}{c}

\NewC{Rn}{\mathbb{R}^n}{real Euclidean $n$-dimensional space}{rsn}
\NewC{GF}{\mathbb{F}_2}{Galois field of size two}{f}

\NewC{N}{\mathcal{N}}{normal distribution}{no}
\NewC{Norm}{\mathcal{N}}{normal distribution}{N}
\NewC{Unif}{\operatorname{Unif}}{uniform distribution}{uz}
\NewF{norm}{||}{||}{Euclidean norm}{0e}
\NewF{Qla}{Q_{\Lambda}(}{)}{nearest neighbor lattice quantizer}{quant}
\NewF{Qco}{Q_{\Code}(}{)}{binary quantizer with respect to a linear code $\Code$}{quant2}
\NewF{E}{\mathbb{E}\left[}{\right]}{expectancy operator}{e}
\NewF{bef}{\operatorname{H}(}{)}{binary entropy function}{he}
\NewF{Pr}{\operatorname{Pr}\left[}{\right]}{probability of an event}{pr}

\newcommand{\IE}{i.e., } 
\newcommand{\EG}{e.g., } 

\newcommand{\vect}[1]{\boldsymbol{#1}}

\newcommand{\D}{\mathrm{d}}

\renewcommand{\epsilon}{\varepsilon}
\renewcommand{\phi}{\varphi}

\makeatletter
\newcommand{\vast}{\bBigg@{5}}
\makeatother

\definecolor{shadecolor}{rgb}{0.97,0.97,0.97}%
\definecolor{framecolor}{rgb}{0,0,0}%

\newcommand{\abbr}[1]{{#1}}				

\makeatletter
\let\aclOLD=\acl
\renewcommand{\acl}[1]{%
  \begingroup    
  \let\@@underline=\relax
  \aclOLD{#1}%
  \endgroup
}
\makeatother

\newcommand{\NewA}[3]{
	\newacronym{#1}{#2}{#3}
}

\NewA{af}{AF}{\abbr{a}mplify-and-\abbr{f}orward}
\NewA{apsk}{APSK}{\abbr{a}mplitude \abbr{p}hase-\abbr{s}hift \abbr{k}eying}
\NewA{ask}{ASK}{\abbr{a}mplitude-\abbr{s}hift \abbr{k}eying}
\NewA{ase}{ASE}{\abbr{a}mplified \abbr{s}pontaneous \abbr{e}mission}
\NewA{awgn}{AWGN}{\abbr{a}dditive \abbr{w}hite \abbr{G}aussian \abbr{n}oise}
\NewA{biawgn}{BI-AWGN}{binary-input \abbr{a}dditive \abbr{w}hite \abbr{G}aussian \abbr{n}oise}
\NewA{bep}{BEP}{\abbr{b}it \abbr{e}rror \abbr{p}robability}
\NewA{ber}{BER}{\abbr{b}it \abbr{e}rror \abbr{r}ate}
\NewA{qap}{QAP}{\abbr{q}uadratic \abbr{a}assignment \abbr{p}roblem}
\NewA{bicm}{BICM}{\abbr{b}it-\abbr{i}nterleaved \abbr{c}oded \abbr{m}odulation}				
\NewA{cm}{CM}{\abbr{c}oded \abbr{m}odulation}				
\NewA{qpsk}{QPSK}{quadrature phase-shift keying}				
\NewA{mlcm}{MLCM}{multilevel coded modulation}				
\NewA{bpsk}{BPSK}{\abbr{b}inary \abbr{p}hase-\abbr{s}hift \abbr{k}eying}				
\NewA{bsc}{BSC}{\abbr{b}inary \abbr{s}ymmetric \abbr{c}hannel}				
\NewA{brgc}{BRGC}{\abbr{b}inary \abbr{r}eflected \abbr{G}ray \abbr{c}ode}				
\NewA{cf}{CF}{\abbr{c}haracteristic \abbr{f}unction}
\NewA{csit}{CSIT}{\abbr{c}annnel \abbr{s}tate \abbr{i}nformation at the \abbr{transmitter}}
\NewA{csi}{CSI}{\abbr{c}annnel \abbr{s}tate \abbr{i}nformation}
\NewA{df}{DF}{\abbr{d}ecode-and-\abbr{f}orward}
\NewA{fd}{FD}{\abbr{f}ull-\abbr{d}uplex}
\NewA{fft}{FFT}{\abbr{f}ast \abbr{F}ourier \abbr{t}ransform}
\NewA{hd}{HD}{\abbr{h}alf-\abbr{d}uplex}
\NewA{iid}{IID}{independent and \abbr{i}dentically \abbr{d}istributed}
\NewA{isi}{ISI}{\abbr{i}nter\abbr{s}ymbol \abbr{i}nterference}
\NewA{lb}{LB}{\abbr{l}ower \abbr{b}ound}
\NewA{map}{MAP}{\abbr{m}aximum \abbr{a} \abbr{p}osteriori}
\NewA{mf}{MF}{\abbr{m}odulo-and-\abbr{f}orward}
\NewA{mlan}{MLAN}{\abbr{m}odulo-\abbr{l}attice \abbr{a}dditive \abbr{n}oise}
\NewA{ml}{ML}{\abbr{m}aximum \abbr{l}ikelihood}
\NewA{mmse}{MMSE}{\abbr{m}inimum \abbr{m}ean \abbr{s}quare \abbr{e}rror}
\NewA{nlpc}{NLPC}{\abbr{n}onlinear \abbr{p}hase \abbr{c}ompensation}
\NewA{nlpn}{NLPN}{\abbr{n}onlinear \abbr{p}hase \abbr{n}oise}
\NewA{nc}{NC}{\abbr{n}etwork \abbr{c}oding}
\NewA{ofmd}{OFDM}{\abbr{o}rthogonal \abbr{f}requency-\abbr{d}ivision \abbr{m}ultiplexing}			
\NewA{dp}{DP}{\abbr{d}ual-\abbr{p}olarization}
\NewA{pam}{PAM}{\abbr{p}ulse \abbr{a}mplitude \abbr{m}odulation}
\NewA{pdf}{PDF}{\abbr{p}robability \abbr{d}ensity \abbr{f}unction}
\NewA{plnc}{PNC}{\abbr{p}hysical-layer \abbr{n}etwork \abbr{c}oding}
\NewA{psk}{PSK}{\abbr{p}hase-\abbr{s}hift \abbr{k}eying}
\NewA{pmd}{PMD}{\abbr{p}olarization \abbr{m}ode \abbr{d}ispersion}
\NewA{pm}{PM}{\abbr{p}olarization-\abbr{m}ultiplexed}
\NewA{pdm}{PDM}{\abbr{p}olarization-\abbr{d}ivision \abbr{m}ultiplexing}
\NewA{qam}{QAM}{\abbr{q}uadrature \abbr{a}mplitude \abbr{m}odulation}
\NewA{sqp}{SQP}{\abbr{s}equential \abbr{q}uadratic \abbr{p}rogramming}
\NewA{rd}{RD}{\abbr{r}adii \abbr{d}istribution}
\NewA{sep}{SEP}{\abbr{s}ymbol \abbr{e}rror \abbr{p}robability}
\NewA{ser}{SER}{\abbr{s}ymbol \abbr{e}rror \abbr{r}ate}
\NewA{si}{SI}{\abbr{s}ide \abbr{i}nformation}
\NewA{sp}{SP}{\abbr{s}ingle-\abbr{p}olarization}
\NewA{sanr}{SNR}{\abbr{s}ignal-to-(additive-)\abbr{n}oise \abbr{r}atio}
\NewA{snr}{SNR}{\abbr{s}ignal-to-\abbr{n}oise \abbr{r}atio}
\NewA{snlse}{sNLSE}{\abbr{s}tochastic \abbr{n}onlinear \abbr{S}chr\"odinger \abbr{e}quation}
\NewA{nlse}{NLSE}{\abbr{n}onlinear \abbr{S}chr\"odinger \abbr{e}quation}
\NewA{stwrc}{sTRC}{\abbr{s}eparated \abbr{t}wo-way \abbr{r}elay \abbr{c}hannel}
\NewA{stwtrc}{sTTRC}{\abbr{s}eparated \abbr{t}wo-way \abbr{t}wo-\abbr{r}elay \abbr{c}hannel}
\NewA{ts}{TS}{\abbr{t}wo-\abbr{s}tage}
\NewA{twrc}{TRC}{\abbr{t}wo-way \abbr{r}elay \abbr{c}hannel}
\NewA{twtrc}{TTRC}{\abbr{t}wo-way \abbr{t}wo-\abbr{r}elay \abbr{c}hannel}
\NewA{wdm}{WDM}{\abbr{w}avelength-\abbr{d}ivision \abbr{m}ultiplexing}
\NewA{scldpc}{SC-LDPC}{spatially coupled low-density parity-check}
\NewA{vn}{VN}{variable node}
\NewA{cn}{CN}{check node}
\NewA{de}{DE}{density evolution}
\NewA{sc}{SC}{spatially-coupled}
\NewA{ldpc}{LDPC}{low-density parity-check}
\NewA{bp}{BP}{belief propagation}
\NewA{dm}{DM}{dispersion-managed}
\NewA{spm}{SPM}{self-phase modulation}
\NewA{xpm}{XPM}{cross-phase modulation}
\NewA{fwm}{FWM}{four-wave-mixing}
\NewA{ixpm}{IXPM}{intrachannel cross-phase modulation}
\NewA{ifwm}{IFWM}{intrachannel four-wave-mixing}
\NewA{ssfm}{SSFM}{split-step Fourier method}
\NewA{fec}{FEC}{forward error correction}
\NewA{psd}{PSD}{power spectral density}
\NewA{ook}{OOK}{on-off keying}
\NewA{smf}{SMF}{single-mode fiber}
\NewA{ssmf}{SSMF}{standard single-mode fiber}
\NewA{dbp}{DBP}{digital backpropagation}
\NewA{edfa}{EDFA}{erbium-doped fiber amplifier}
\NewA{ofdm}{OFDM}{orthogonal frequency division multiplexing}
\NewA{exit}{EXIT}{extrinsic information transfer}
\NewA{pexit}{P-EXIT}{protograph extrinsic information transfer}
\NewA{osnr}{OSNR}{optical signal-to-noise ratio}
\NewA{roadm}{ROADM}{reconfigurable optical add-drop multiplexer}
\NewA{rps}{RPS}{Raman pump station}
\NewA{mlse}{MLSE}{maximum likelihood sequence estimation}
\NewA{dcf}{DCF}{dispersion compensating fiber}
\NewA{tcm}{TCM}{trellis coded modulation}
\NewA{edc}{EDC}{electronic dispersion compensation}
\NewA{ldpcc}{LDPCC}{low-density parity-check convolutional}
\NewA{llr}{LLR}{log-likelihood ratio}
\NewA{ra}{RA}{repeat-accumulate}
\NewA{ira}{IRA}{irregular-repeat-accumulate}
\NewA{ara}{ARA}{accumulate-repeat-accumulate}
\NewA{mi}{MI}{mutual information}
\NewA{vdmm}{VDMM}{variable degree matched mapping}
\NewA{gmi}{GMI}{generalized mutual information}
\NewA{wd}{WD}{windowed decoder}
\NewA{gn}{GN}{Gaussian noise}

\newacronym[%
	longplural={binary erasure channels},%
	shortplural={BECs}%
]{bec}{BEC}{binary erasure channel}%

\renewcommand{\imag}{\ensuremath{j}}
\renewcommand{\vect}[1]{\ensuremath{\mathbf{#1}}}

\newcommand{\VNs}{\glspl{vn}\xspace}
\newcommand{\CNs}{\glspl{cn}\xspace}
\newcommand{\CnNum}{c}

\newcommand{\precision}{\delta}
\newcommand{\pe}{p_{\text{tar}}}

\newcommand{\lsuccess}{l_\text{s}(\mathbf{A}, \rho)}
\newcommand{\ls}{l_\text{s}}

\newcommand{\Aunif}{\mathbf{A}_\text{uni}}
\newcommand{\Aopt}{\ensuremath{\mathbf{A\!{ }^*}}}

\newcommand{\conv}{*}
\newcommand{\transpose}{\intercal}
\newcommand{\hermitian}{\dagger}

\newcommand{\Lsp}{\ensuremath{L_{\text{sp}}}}
\newcommand{\Nsp}{\ensuremath{N_\text{sp}}}
\newcommand{\SEF}{\ensuremath{n_\text{sp}}}

\newcommand{\PSDl}{\ensuremath{{N}_{\text{EDFA}}}}

\newcommand{\PolX}{\ensuremath{\mathsf{x}}}
\newcommand{\PolY}{\ensuremath{\mathsf{y}}}

\newcommand{\LF}{\ensuremath{{M}}} 
\newcommand{\TL}{\ensuremath{{T}}} 
\newcommand{\WS}{\ensuremath{{W}}} 

\newcommand{\Pase}{\ensuremath{P_{\text{ASE}}}}

\newcommand{\Barja}[1]{\ensuremath{\proto^{(#1)}}}

\newcommand{\Carja}[1]{\ensuremath{\mathcal{C}_{\text{AR4JA}#1}}}
\newcommand{\Csc}[1]{\ensuremath{\mathcal{C}_{\text{SC}#1}}}

\newcommand{\proto}{\ensuremath{\mathbf{P}}}

\definecolor{DarkGreen}{rgb}{0.0, 0.5, 0.0}

\newcommand{\RevA}[1]{{#1}}
\newcommand{\RevB}[1]{{#1}}
\newcommand{\RevC}[1]{{#1}}

\newenvironment{DIFnomarkup}{}{}

\addto\captionsenglish{}

\begin{document}

\begin{DIFnomarkup}

\end{DIFnomarkup}

\title{Improving soft FEC performance for higher-order modulations
\RevC{via optimized bit channel mappings}}



\newcommand{\aff}[1]{${ }^{#1}$}

\author{%
	Christian H\"ager,\aff{1,\ast} 
	Alexandre Graell i Amat,\aff{1}
	Fredrik Br\"annstr\"om,\aff{1} 
	Alex Alvarado,\aff{2} and
	Erik Agrell\aff{1}}

\address{%
	\aff{1}Department of Signals and Systems, Chalmers University of Technology, Gothenburg, Sweden \\ 
	\aff{2}Optical Networks Group, Department of Electronic and Electrical Engineering, University College London, London WC1E 7JE, UK\\
	\aff{\ast}{\color{blue}christian.haeger@chalmers.se}
	}%
%

\begin{abstract}
	Soft \acl{fec} with higher-order modulations is often implemented in
	practice via the pragmatic \acl{bicm} paradigm, where a single
	binary code is mapped to a nonbinary modulation. In this paper, we
	study the optimization of the mapping of the coded bits to the
	modulation bits for a polarization-multiplexed fiber-optical system
	without optical inline dispersion compensation. Our focus is on
	protograph-based \gls{ldpc} codes which allow for an efficient
	hardware implementation, suitable for high-speed optical
	communications. The optimization is applied to the AR4JA protograph
	family, and further extended to protograph-based spatially coupled
	LDPC codes assuming a \acl{wd}. Full field simulations via the
	\acl{ssfm} are used to verify the analysis. The results show
	performance gains of up to 0.25 dB, which translate into a possible
	extension of the transmission reach by roughly up to 8\%, without
	significantly increasing the system complexity. 
\end{abstract}



\renewcommand{\refname}{References and links}

\bibliographystyle{osajnl}

\glsresetall

\section{Introduction}



There is currently a large interest in developing practical \gls{cm}
schemes that can achieve high spectral efficiency close to the
ultimate capacity limits of optical fibers \cite{Essiambre2010}. 
Pragmatic \gls{bicm} in combination with \gls{ldpc} codes is one of
the most popular capacity-approaching CM techniques for achieving high
spectral efficiency, due to its simplicity and flexibility
\cite{Costello2007}. For a \gls{bicm} system, a helpful abstraction is
to think about transmitting data using a single \gls{fec} encoder over
a set of parallel binary-input channels, or simply bit channels, with
different qualities. This is due to the fact that bits are not
protected equally throughout the signal constellation. With this
useful picture, an immediate problem is how to best allocate the coded
bits from the encoder to these channels. As a baseline, a random or
consecutive/sequential mapping is commonly used in practice. \RevC{
However, by optimizing the mapping strategy, one can improve the
system performance, at almost no increased complexity cost. While
\gls{bicm} has been studied for fiber-optical communications by many
authors, see \EG \cite{Smith2012} or \cite{Djordjevic2009} and
references therein, to the best of our knowledge, optimized bit
channel mappings have not yet been studied for such systems. In the
following, we use the term ``bit mapper'' to denote the device that
performs the bit channel mapping. We remark that other terms, \EG
``bit interleaver'' or ``mapping device'', are also frequently used in
the literature. } 

In this paper, we address the bit mapper optimization for a BICM
system based on LDPC codes in the context of long-haul fiber-optical
communications. Our target system operates over a communication link
with a lumped amplification scheme and without optical inline
dispersion compensation. In general, the signal undergoes a
complicated evolution and interacts with \gls{ase} noise and
co-propagating signals through dispersive and nonlinear effects. For
dispersion uncompensated transmission, it has been shown that an
additive \gls{gn} model can be assumed, provided that dispersive
effects are dominant and nonlinear effects are weak \cite{Beygi2012,
Carena2012}. We use the \gls{gn} model for our analysis, which
accounts for both the \gls{ase} noise from inline \glspl{edfa} and
nonlinear noise due to the optical Kerr effect.

The starting point for the optimization problem is a fixed modulation
format and a given error correction code, \IE we do not consider the
\emph{joint} design of the modulation, bit mapper, and code. This
scenario is often encountered in practice when the modulation and code
have been designed separately and/or are predetermined according to
some communication standard. Our focus is on protograph-based
\gls{ldpc} codes \cite{Thorpe2005}, which are very attractive from a
design perspective and allow for a high-speed hardware implementation,
suitable for fiber-optical communications \cite{Schmalen2013}. A
protograph is a (small) bipartite graph, from which the Tanner graph
defining the code is obtained by a copy-and-permute procedure.  As one
illustrative example for protograph-based codes, we consider the AR4JA
protographs developed by researchers from JPL/NASA in
\cite{Divsalar2005}. We also consider bit mapper optimization for
protograph-based \gls{scldpc} codes using the \gls{wd} proposed in
\cite{Iyengar2012}. \gls{scldpc} codes, originally introduced as
\gls{ldpc} convolutional codes in \cite{JimenezFelstrom1999}, have attracted a lot
of attention due to their capacity-achieving performance under
\gls{bp} decoding for a variety of communication channels
\cite{Kudekar2011}. \gls{scldpc} codes can be constructed using
protographs and they are considered as viable candidates for future
spectrally efficient fiber-optical systems \cite{Schmalen2013}.

Most of the literature about bit mapper optimization deals with
irregular \gls{ldpc} codes that are not based on protographs, see \EG
\cite{Cheng2012, Richter2007}. Attempts to improve the performance of
\gls{bicm} systems with protograph-based codes through bit mapper
optimization have been previously made in \cite{Divsalar2005a,Jin2010,
Nosratinia2011a}. In \cite{Divsalar2005a}, a mapping strategy inspired
by the waterfilling algorithm for parallel channels called \gls{vdmm}
is presented. This idea is extended in \cite{Jin2010}, where the authors
exhaustively search over all possible nonequivalent connections
between protograph nodes and modulation bits showing performance
improvements over \gls{vdmm}. As pointed out in
\cite{Nosratinia2011a}, the above approaches are somewhat restrictive
in the sense that only certain protographs can be used with certain
modulation formats. A more flexible approach is proposed in
\cite{Nosratinia2011a}, which is in principle suitable for any
protograph structure and modulation but relies on a larger
intermediate protograph.

Our optimization of the bit mapper is based on the decoding threshold
over the \gls{awgn} channel \RevA{similar to, \EG \cite{Cheng2012,
Richter2007, Jin2010},} albeit assuming a fixed number of decoding
iterations. The decoding threshold divides the channel quality
parameter range (in our case the equivalent \gls{snr} of the \gls{gn}
model) into a region where reliable decoding is possible and where it
is not. In the asymptotic case, \IE assuming infinite codeword length,
\gls{de} or one-dimensional simplifications via \gls{exit} functions
can be used to find the decoding threshold for \gls{ldpc} codes under
\gls{bp} decoding \cite{Richardson2001}. Approximate decoding
thresholds of protograph-based codes assuming binary modulation can be
obtained by using the \gls{pexit} analysis \cite{Liva2007}. \RevA{The
approach proposed here relies on a modified \gls{pexit} analysis which
allows for a fractional allocation between protograph nodes and
modulation bits. This approach is, to the best of our knowledge, novel
in the context of protograph-based codes and different from the
approaches described in \cite{Divsalar2005a,Jin2010, Nosratinia2011a}.
In particular, a fractional allocation} allows for an unrestricted
matching of protographs and modulation formats and additionally does
not suffer from an increased design complexity due to a larger
intermediate protograph. We also discuss several ways to reduce the
optimization complexity. \RevA{In particular, we introduce periodic
bit mappers for SC-LDPC codes} with a WD, which is based on the
results we previously presented in \cite{Hager2014}, where optimized
bit mappers are found for (nonprotograph-based) SC-LDPC codes assuming
parallel \glspl{bec} without considering the WD. The use of a \gls{wd}
in this paper is motivated by the reduced complexity and decoding
delay with respect to full decoding.  Finally, we provide a simulative
verification assuming both linear and nonlinear transmission
scenarios. For the latter case, we use the \gls{ssfm} to show that the
performance improvements predicted from the \gls{awgn} analysis can be
achieved for a realistic transmission scenario including nonlinear
effects.

\subsection{Notation}

Vectors and matrices are typeset in bold font by lowercase letters
$\mathbf{a}$ and capital letters $\mathbf{A}$, respectively. Matrix
transpose is denoted by $(\cdot)^\transpose$, Hermitian transpose by
$(\cdot)^\hermitian$, and the squared norm of a complex vector by
$\|\vect{a}\|^2$.  $\vect{I}_n$ denotes the identity matrix of size
$n$. Complex conjugation is denoted by $(\cdot)^{*}$. $\delta(t)$ is
Dirac's delta function, whereas $\delta[k]$ is the Kronecker delta.
Convolution is denoted by $\conv$. $\mathbb{N}_0$, $\mathbb{R}$, and
$\mathbb{C}$ denote the set of nonnegative integers, real numbers, and
complex numbers, respectively. Random variables and vectors are
denoted by capital letters and their realizations by lowercase
letters. The \gls{pdf} of a random variable $Y$ conditioned on the
realization of another random variable $X$ is denoted by
$f_{Y|X}(y|x)$, and the expected value by $\mathbb{E}[\cdot]$.

\section{System model}
\label{sec:SystemModel}

\subsection{Continuous-time channel}

\begin{figure}[t]
	\centering
		\includegraphics{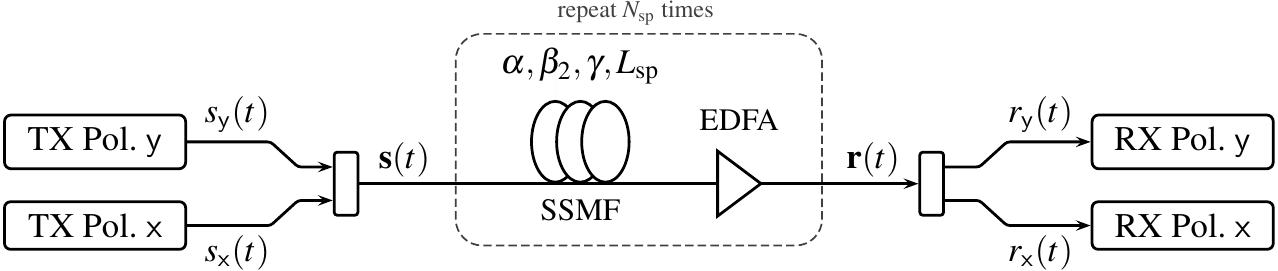}
	\caption{Block diagram of the consider fiber-optical transmission
	system.}
	\label{fig:BlockDiagram}
\end{figure}

We consider transmission of a \gls{pm} signal over a \gls{ssmf} with a
lumped amplification scheme as shown in Fig.~\ref{fig:BlockDiagram}.
The optical link consists of $\Nsp$ spans of \gls{ssmf} with length
$\Lsp$. The baseband signal in each polarization is generated via a
linear pulse modulation according to $s_{\PolX}(t) = \sum_k s_{\PolX,
k} p(t-k/R_s)$, where $s_{\PolX, k} \in \mathbb{C}$ are the
information symbols, $p(t)$ the real-valued pulse shape, and $R_s$ the
symbol rate. (We give expressions for polarization $\PolX$ only, if
polarization $\PolY$ has an equivalent expression.) The \gls{pm}
signal $\vect{s}(t) = (s_\PolX(t), s_\PolY(t))^\transpose$ is launched
into the fiber and propagates according to \cite[Ch.~3]{Agrawal2005}
\begin{align}
	\frac{\partial \vect{v}(t,z) }{\partial z}
	= - \frac{\alpha - g(z)}{2} \vect{v}(t,z) - \imag \frac{\beta_2}{2}
	\frac{\partial^2 \vect{v}(t,z) }{\partial t^2}
	+ \imag \gamma \vect{v}(t,z) \|\vect{v}(t,z)\|^2 + \vect{w}(t,z), 
	\label{eq:manakov}
\end{align}
where $\vect{v}(t,z)$ is the complex baseband representation of the
electric field and the input to the first fiber span and the output
signal are $\vect{s}(t) = \vect{v}(t,0)$ and $\vect{r}(t) =
\vect{v}(t, \Nsp \Lsp)$, respectively. In \eqref{eq:manakov}, $\alpha$
is the attenuation coefficient, $\beta_2$ the chromatic dispersion
coefficient, and $\gamma$ the nonlinear Kerr parameter. The terms
$g(z)$ and $\vect{w}(t,z) = (w_\PolX(t,z),
w_\PolY(t,z))^\transpose$ model the amplifier gain and the generated
\gls{ase} noise \cite[p.~84]{Secondini2008}. Each \gls{edfa}
introduces circularly symmetric complex Gaussian noise with two-sided
\gls{psd} $\PSDl = (G - 1) h \nu_s \SEF$
\cite[eq.~(54)]{Essiambre2010} per polarization, where $G = e^{\alpha
\Lsp}$ is the amplifier gain, $h$ is Planck's constant, $\nu_s$ the
carrier frequency, and $\SEF$ the spontaneous emission factor.  A
standard coherent linear receiver is used, consisting of an equalizer,
a pulse-matched filter and a symbol-time sampler. This amounts to
$r_{{\PolX}, k} = \left.  r_{\PolX}(t) \conv h(t) \conv p(-t)
\right|_{t = k / R_s}$, where the frequency response of the equalizer
$h(t)$ is $H(f) = \exp(\imag 2 \beta_2 \pi^2 f^2 \Nsp \Lsp)$.

\subsection{Discrete-time channel}

An approximate discrete-time model for the received samples
$\bold{r}_k = (r_{\PolX, k}, r_{\PolY, k})^\transpose$ based on the
transmitted symbols $\bold{s}_k = (s_{\PolX, k}, s_{\PolY,
k})^\transpose$  is given by $\vect{r}_{k} \approx \zeta \vect{s}_{k}
+ \vect{n}_{k} + \tilde{\vect{n}}_{k}$, where $\zeta \in \mathbb{C}$
\cite{Beygi2012}.  The term $\bold{n}_k = (n_{\PolX, k}, n_{\PolY,
k})^\transpose$ accounts for the linear \gls{ase} noise with
$\mathbb{E}[\vect{N}_{k} \vect{N}^\hermitian_{k'}] = \Pase \vect{I}_2
\delta[k-k']$, where $\Pase = \Nsp \PSDl R_s $. The term
$\tilde{\bold{n}}_k = (\tilde{n}_{\PolX, k}, \tilde{n}_{\PolY,
k})^\transpose$ accounts for nonlinear noise with
$\mathbb{E}[\tilde{\vect{N}}_{k} \tilde{\vect{N}}^\hermitian_{k'}] =
\eta P^3 \vect{I}_2 \delta[k-k']$, where $P =\lim_{T \to \infty}
(\int_{-T}^{T} s_{\PolX}(t)^2 \, \D t)/(2T)$ is the transmit power per
polarization (assumed to be equal for both polarizations). $\eta$ is a
function of the link parameters $\alpha, \beta_2, \gamma,\Lsp, \Nsp$
and the symbol rate $R_s$ \cite[eq. (15)]{Beygi2012}, and $|\zeta|^2 =
1 - |\eta| P^2$. The conditional \gls{pdf} in this model is assumed to
be Gaussian according to
\begin{equation}
		f_{\vect{R}_k|\vect{S}_k}(\vect{r}_k|\vect{s}_k) =
		\frac{1}{\left( \pi P_{\text{N}} \right)^2} 
		\exp \left(
			- \frac{\|\vect{r}_k - \zeta
			\vect{s}_k \|^2}{P_{\text{N}} }
		\right),
    \label{eq:gauss}
\end{equation}
where $P_{\text{N}} = \Pase + \eta P^3$. The equivalent \gls{snr} is
defined as $\rho \triangleq |\zeta|^2 P / ( \Pase + \eta P^3)$.

\subsection{Bit-interleaved coded modulation}

\begin{figure}[t]
	\centering
		\subfloat[]{\includegraphics{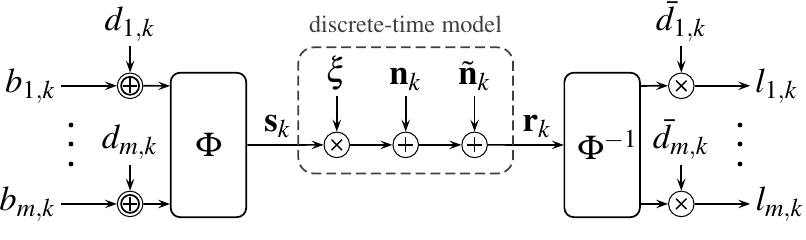}}
		\qquad
		\subfloat[]{\includegraphics{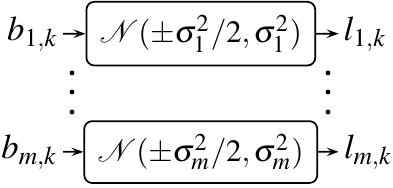}}
	\caption{(a) BICM block diagram including the channel symmetrization
	technique. (b) Approximate model with parallel Gaussian LLR
	channels. }
	\label{fig:bit_channels}
\end{figure}

\begin{figure}[t]
	\centering
		\subfloat[]{\includegraphics{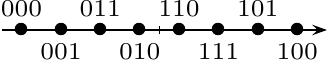}}
		\qquad
		\subfloat[]{\includegraphics{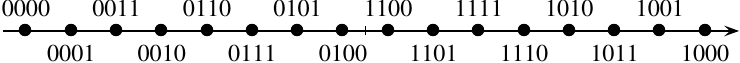}}
	\caption{The considered signal constellations in each dimension.}
	\label{fig:constellations}
\end{figure}

The transmitted symbols $\vect{s}_k$ in each time instant $k$ take on
values from a discrete signal constellation $\mathcal{X} \subset
\mathbb{C}^2$. Each point in the constellation is labeled with a
unique binary string of length $m = \log_2 |\mathcal{X}|$, where
$b_i(\vect{a})$, $1\leq i \leq m$, denotes the $i$th bit in the binary
string assigned to $\vect{a} \in \mathcal{X}$ \RevC{(counting from
left to right)}. Consider now the block diagram shown in
Fig.~\ref{fig:bit_channels}(a), \RevB{where the modulo 2 addition of
$d_{i,k}$ and multiplication by $\bar{d}_{i,k} = (-1)^{d_{i,k}}$ is
explained further below and can be ignored for now.} At each time
instant $k$, the modulator $\Phi$ takes $m$ bits $b_{i,k}$, $1\leq
i\leq m$, and maps them to one of the constellation points according
to the binary labeling. We consider two product constellations of
one-dimensional constellations labeled with the \gls{brgc} as shown in
Fig.~\ref{fig:constellations}, which we refer to as PM-$64$-QAM and
PM-$256$-QAM.
At the receiver, the demodulator $\Phi^{-1}$ computes soft reliability
information about the transmitted bits in the form of the
\glspl{llr}
\begin{equation}
	l_{i,k} \triangleq \log
	\frac{f_{\vect{R}_k|B_{i,k}}(\vect{r}_{k} | 0)}
	{f_{\vect{R}_{k}|B_{i,k}}(\vect{r}_{k} | 1)} = 
	\log 
		\frac{\sum_{\vect{s} \in \setX_{i, 0}} 
		f_{\vect{R}_k|\vect{S}_k}(\vect{r}_k|\vect{s})}
		{\sum_{\vect{s} \in \setX_{i, 1}}
		f_{\vect{R}_k|\vect{S}_k}(\vect{r}_k|\vect{s})}, 
    \label{eq:Lvalue_exact}
\end{equation}
where $\mathcal{X}_{i,u} \triangleq \{\vect{a} \in \mathcal{X}:
b_i(\vect{a}) = u\}$ is the subconstellation where all
points have the bit $u$ at the $i$th position of their binary label.

A useful way to think about the setup depicted in
Fig.~\ref{fig:bit_channels}(a) is to imagine transmitting over a set
of parallel bit channels, where one may interpret the conditional
distribution of the \gls{llr} $f_{L_{i,k}|B_{i,k}}(\cdot|\cdot)$ as a
bit channel. In the following, we say that a bit channel
$f_{L|B}(l|b)$ is symmetric if $f_{L|B}(l|0) = f_{L|B}(-l|1)$ and the
channel is referred to as an LLR channel if $f_{L|B}(l|0) e^l =
f_{L|B}(l|1)$.  \RevB{One can show that
$f_{L_{i,k}|B_{i,k}}(\cdot|\cdot)$ is an LLR channel, but not
necessarily symmetric in general. Symmetry can be enforced by adding
modulo 2 independent and identically distributed bits $d_{i,k}$ to the
bits $b_{i,k}$ and multiplying the corresponding LLR by
$\bar{d}_{i,k}$  (see Fig.~\ref{fig:bit_channels}(a)) \cite{Hou2003}.
The symmetry condition is an important requirement for the analysis in
Section~\ref{sec:exit}, where one implicitly relies on the assumption
that the all-zero codeword has been transmitted \cite[p.~389]{Ryan2009}.}

\begin{figure}[t]
	\centering
		\subfloat[$\rho = 10$ dB]{\includegraphics{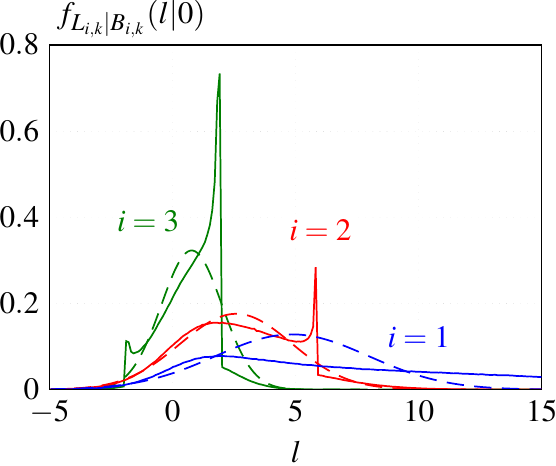}}
		\qquad
		\subfloat[$\rho = 17$ dB]{\includegraphics{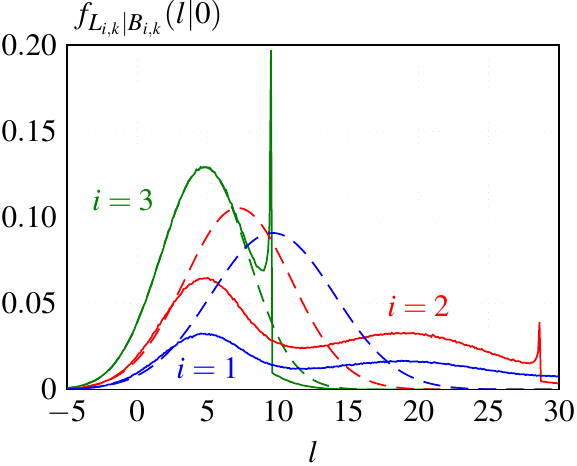}}
	\caption{\RevC{Comparison of the LLR channels for PM-$64$-QAM including channel
	symmetrization (solid lines) with the Gaussian LLR channels that have
	the same MI (dashed lines).}}
	\label{fig:llrs_snr}
\end{figure}

To simplify the analysis, the original bit channels are replaced with
parallel symmetric Gaussian LLR channels, as shown in
Fig.~\ref{fig:bit_channels}(b), where an LLR channel $f_{L|B}(l|b)$ is
called a symmetric Gaussian LLR channel with parameter $\sigma^2$ if
$L \sim \mathcal{N}(\sigma^2/2, \sigma^2)$ conditioned on $B = 0$ and
$L \sim \mathcal{N}(-\sigma^2/2, \sigma^2)$ conditioned on $B = 1$.
In order to find a correspondence between the LLR channels
$f_{L_{i,k}|B_{i,k}}(\cdot|\cdot)$ and the parameters $\sigma^2_i$,
one may match the \gls{mi} according to $\sigma_i^2 =
J^{-1}(I_i(\rho))^2$, where $I_i(\rho) = I(B_{i,k};L_{i,k})$ is
independent of $k$ and $J(\sigma)$ denotes the \gls{mi} between the
output of a symmetric Gaussian LLR channel and uniform input bits.
\RevC{As an example and to visualize the different bit channel
qualities, in Fig.~\ref{fig:llrs_snr} we compare the LLR channels
(solid lines, estimated via histograms) with the approximated Gaussian
LLR channels (dashed lines) assuming an \gls{awgn} channel and two
different values of $\rho$ for the three distinct bit channels of
PM-$64$-QAM (see Fig.~\ref{fig:constellations}(a)). It can be seen
that the actual densities are clearly non-Gaussian. However, the
Gaussian approximation is quite accurate for the bit mapper
optimization as shown later and allows for a major simplification of
the analysis, thereby justifying its use. }

\begin{figure}[t]
	\centering
		\includegraphics{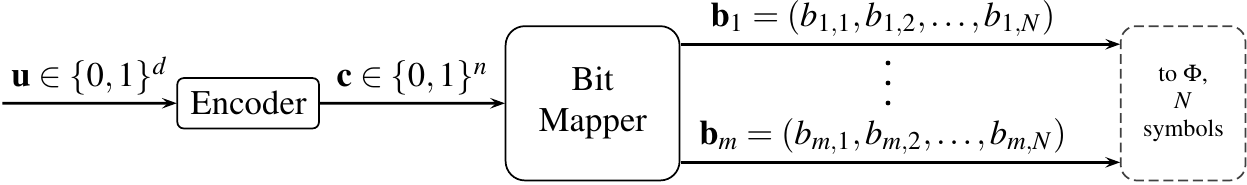}
	\caption{Block diagram illustrating the purpose of the bit mapper. }
	\label{fig:bit_mapper}
\end{figure}

Consider now the case where a binary code $\mathcal{C} \subset
\{0,1\}^n$ of length $n$ and dimension $d$ is employed and each
codeword $\vect{c} = (c_1, \dots, c_n)$ is transmitted using $N = n/m$
symbols $\vect{s}_k$. The allocation of the coded bits to the
modulation bits (\IE the different bit channels in
Fig.~\ref{fig:bit_channels}(b)) is determined by a bit mapper as shown
in Fig.~\ref{fig:bit_mapper}, where the vectors $\vect{b}_1$, \ldots,
$\vect{b}_m$ are of length $N$. Our goal is to find good bit mappers
for a fixed code and modulation. As a baseline, we consider a
consecutive mapper according to $b_{i,k} = c_{(k-1)m+i}$ for $1 \leq i
\leq m$, $1 \leq k \leq N$.

\section{Protograph-based LDPC codes}
\label{sec:Protograph}

An \gls{ldpc} code of length $n$ and dimension $d$ is defined via a
sparse parity-check matrix $\bold{H} = [h_{i,j}] \in \{0,1\}^{\CnNum
\times n}$, where $\CnNum = n-d$. There exist different methods to
construct ``good'' \gls{ldpc} codes, \IE good matrices $\bold{H}$.
One popular method is by using protographs \cite{Thorpe2005}.  An
\gls{ldpc} code can be represented by using a bipartite Tanner graph
consisting of $n$ \VNs and $\CnNum$ \CNs, where the $i$th \gls{cn} is
connected to the $j$th \gls{vn} if $h_{i,j} = 1$. A protograph is also
a bipartite graph defined by an adjacency matrix $\proto = [p_{i,j}]
\in \mathbb{N}_0^{\CnNum' \times n'}$, called the base matrix. Given
$\proto$, a parity-check matrix $\vect{H}$ is obtained by replacing
each entry $p_{i,j}$ in $\proto$ with a random binary $\LF$-by-$\LF$
matrix which contains $p_{i,j}$ ones in each row and column. This
procedure is called lifting and $\LF \geq \max_{i,j} p_{i,j}$ is the
so-called lifting factor. Graphically, this construction amounts to
copying the protograph $\LF$ times and subsequently permuting edges.
Parallel edges, \IE for $p_{i,j}>1$, are permitted in the protograph
and are resolved in the lifting procedure.  The design rate of the
code is given by $R = 1 - \CnNum/n = 1 - \CnNum'/n'$, where $c = c'
\LF$ and $n = n' \LF$.

\subsection{AR4JA codes}

As one example to illustrate the bit mapper optimization technique, we
consider the AR4JA code family defined by the protographs in
\cite[Fig.~8]{Divsalar2005}. The base matrix $\Barja{\ell}$ of the
AR4JA code ensemble with parameter $\ell \in \mathbb{N}_0$ can be
recursively defined via \cite{Nosratinia2011a}
\begin{align}
	\Barja{\ell} = 
	\left(\!\!\!
	\begin{array}{c | c}
		\Barja{\ell -1} &
		\begin{matrix}
			 0 & 0 \\
			 3 & 1 \\
			 1 & 3 
		\end{matrix}
	\end{array} 
	\!\!\!
	\right), 
	\qquad
  \Barja{\ell = 0} = 
	\begin{pmatrix}
		1 & 2 & 0 & 0 & 0 \\
		0 & 3 & 1 & 1 & 1 \\
		0 & 1 & 2 & 2 & 1 
	\end{pmatrix} 
\end{align}
with $c' = 3$ and $n' = 5 + 2\ell$. \VNs corresponding to the second
column of the base matrix are punctured, leading to a design rate of
$R = (1-c'/n') \cdot n'/(n'-1) = (\ell+1)/(\ell+2)$. 

\subsection{Spatially coupled LDPC codes}
\label{sec:scldpc}

\gls{scldpc} codes have parity-check matrices with a band-diagonal
structure (for a general definition see, \EG \cite{Kudekar2011}).
For completeness, we briefly review the construction via protographs
in \cite{Mitchell2011}, \cite[Sec.~II-B]{Iyengar2012}. The base matrix
$\proto_{[\TL]}$ of a $(J,K)$ regular, protograph-based \gls{scldpc}
code with termination length $\TL$ can be constructed by specifying
matrices $\proto_i$, $0 \leq i \leq m_\text{s}$ of dimension $J'$
by $K'$, where $m_\text{s}$ is referred to as the memory. The matrices
are such that $\proto = \sum_{i=0}^{m_\text{s}} \proto_i$ has
column weight $J$ and row weight $K$ for all columns and rows,
respectively. Given $\TL$ and the matrices $\proto_i$, the base
matrix $\proto_{[\TL]}$ is constructed as

\newcommand\overmat[2]{%
  \makebox[0pt][l]{$\smash{\color{white}\overbrace{\phantom{%
    \begin{matrix}#2\end{matrix}}}^{\text{\color{black}#1}}}$}#2}

\newcommand\bovermat[2]{%
  \makebox[0pt][l]{$\smash{\overbrace{\phantom{%
    \begin{matrix}#2 & & a \end{matrix}}}^{\text{#1}}}$}#2}

\begin{align}
	\proto_{[\TL]} = 
	\begin{pmatrix}
		\bovermat{\RevB{$\TL$ times}}{\proto_0 &  &  } \\ 	
		\proto_1 & \ddots & \\	
		\vdots &  \ddots & \proto_0 \\	
		\proto_{m_s} & \ddots & \proto_1 \\	
		 & \ddots & \vdots  \\	
		 & & \proto_{m_s}\\	
	\end{pmatrix}.
\end{align}
From the dimensions of $\proto_{[\TL]}$ one can infer a design rate of
$R(\TL) = 1 - ( \TL + m_\text{s}) J' / (\TL K')$. As $\TL$ grows
large, the rate approaches $R(\infty) = 1 - J'/K'$.

Since our goal is not to optimize the code, we rely on base matrices
that have been proposed elsewhere in the literature, in particular in
combination with a \gls{wd} which we discuss below. We consider
$\proto_0 = (2, 2, 2)$ and $\proto_1 = (1, 1, 1)$ according to
\cite[Design rule 1]{Iyengar2012}, where $J'=1$, $K'=3$, $m_\text{s} =
1$, and $R(\infty) = 2/3$.

\subsection{Decoding and asymptotic EXIT analysis}

\label{sec:exit}

We use a modified version of the \gls{pexit} analysis as a tool to
predict the iterative \gls{bp} performance behavior of the
protograph-based codes \cite{Liva2007}. A detailed description of this
tool for binary modulation is available in \cite{Liva2007} and
\cite[Algorithm 9.2]{Ryan2009}. Here, we only describe the necessary
modifications to account for the \gls{wd} and the nonbinary
modulations. We start with the former and explain the latter in the
next section.

We employ the \gls{wd} scheme developed in \cite{Iyengar2012}.
\gls{wd} helps to alleviate the long decoding delays and high decoding
complexity of \gls{scldpc} codes under full \gls{bp} decoding by
exploiting the fact that two \VNs are not involved in the same
parity-check equation if they are at least $(m_\text{s} + 1)K'$
columns apart \cite{Iyengar2012}. The \gls{wd} restricts message
updates to a subset of \VNs and \CNs in the entire graph. After a
predetermined number of decoding iterations, this subset changes and
the decoding window slides to the next position. Pseudocode for the
modified \gls{pexit} analysis of \gls{scldpc} codes accounting for the
\gls{wd} is presented in Algorithm 1. The main difference with respect
to \gls{bp} decoding is the window size parameter $\WS$, which
specifies the number of active \CNs in the protograph considered in
each window as a multiple of $J'$. The \gls{pexit} analysis for the
standard \gls{bp} decoder can be recovered from Algorithm 1 by setting
$\TL = 1$, $\WS = 1$, $J'=c'$, and $K'=n'$.

\IncMargin{1em}
\newcommand{\maxIter}{\ensuremath{l_\text{max}}}
\newcommand{\CnStart}{\ensuremath{c_\text{start}}}
\newcommand{\CnEnd}{\ensuremath{c_\text{end}}}
\newcommand{\VnStart}{\ensuremath{v_\text{start}}}
\newcommand{\VnEnd}{\ensuremath{v_\text{end}}}
\newcommand{\TnStart}{\ensuremath{t_\text{start}}}
\newcommand{\TnEnd}{\ensuremath{t_\text{end}}}
\begin{algorithm}[t]
	\small
	\SetKw{ShortFor}{for}
	\SetKw{Break}{break}
	\SetKw{MyWhile}{while}
	\SetKw{MyIf}{if}
	\SetKw{MySet}{set}
	\SetKw{MyElse}{else}
	\SetKw{MyCompute}{compute}
	\SetKwFunction{VNupdate}{VNupdate}
	\SetKwFunction{CNupdate}{CNupdate}
	\SetKwFunction{VNinit}{VNinit}
	\SetKwFunction{VN}{VN}
	\SetKwFunction{mean}{mean}
	\SetKwFunction{errProb}{error probability}
	\SetKwFunction{OutMsg}{Messages}

	\KwIn{$\maxIter$ (max.~iterations per window), $\pe$ (target error
	probability), $\WS$, ($J', K'$), $\rho$} 
	\KwOut{$S$ (decoding success, either true or false), $\ls$
	(iterations until successful decoding)}
	\BlankLine
	\For(\tcc*[f]{initialization of channel variances for VNs}){$i=1$ \KwTo $n'$}{
		\MyIf \VN $i$ is punctured \MySet $\sigma^2_i = 0$\tcc*[r]{treat as an erasure}
		\MyElse \MySet $\sigma^2_i = f(\rho)$\tcc*[r]{E.g.,$f(\rho) = 8 R \rho$ if $\rho = E_b/N_o$ \cite{Ryan2009}}
	}
	$\ls = 0$ \tcc*[r]{total iteration counter}
	\For{$j = -\WS+2$ \KwTo $\TL$}{
	$\CnStart \leftarrow \max((j-1)J' + 1, 1)$ \tcc*[r]{first index of active CNs}
	$\CnEnd   \leftarrow \min((\WS+j-1)J', m)$ \tcc*[r]{last index of active CNs}
	$\VnStart \leftarrow \max((j-1)K' + 1, 1)$\tcc*[r]{first index of active VNs}
	$\VnEnd   \leftarrow \min((\WS+j-1)K', n')$\tcc*[r]{last index of active VNs}
	$\TnStart \leftarrow \max((j-1)K' + 1, 1)$\tcc*[r]{first index of target VNs}
	$\TnEnd   \leftarrow \max((j-1)K' +K', K')$\tcc*[r]{last index of target VNs}
	$l = 0$\; 
	\While{$l \leq \maxIter$}{
		\MyIf \mean(error probability of \VN \TnStart\, to \TnEnd) $< \pe$ \Break
		\MyWhile\;
		\ShortFor $i=\VnStart$ \KwTo $\VnEnd$ \MyCompute \OutMsg($\sigma_i^2$) of VN $i$
		\tcc*[r]{Eq. (9.46) \cite{Ryan2009}}
		\ShortFor $i=\CnStart$ \KwTo $\CnEnd$ \MyCompute \OutMsg of CN $i$ 
		\tcc*[r]{Eq. (9.47) \cite{Ryan2009}}
	  $l \leftarrow l +1$ and $\ls \leftarrow \ls +1$\;
	}
	}
	\MyIf \mean(error probability of \VN $1$\, to $n'$) $< \pe$ \MySet $S = 1$
	\MyElse \MySet $S = 0$
	\caption{\gls{pexit} analysis of the \gls{wd} for a ($J, K$) regular
	\gls{scldpc} protograph.  }
\end{algorithm}
\DecMargin{1em}


\section{Bit mapper optimization}
\label{sec:Optimization}

\subsection{Asymptotic bit mapper model}

Each VN in the protograph represents $\LF$ \VNs in the lifted Tanner
graph. Since a VN corresponds to one bit in a codeword, the $n'$ \VNs
in the protograph give rise to $n'$ different classes of coded bits
that are treated as statistically equivalent in the \gls{pexit} analysis.
In particular, for binary modulation, each protograph VN is assigned
with one input variance, corresponding to either a punctured bit or
the Gaussian LLR channel (see lines $2$ and $3$ in Algorithm 1). For
nonbinary modulations, \VNs in the same class can in principle have
different input densities. Assume for example that a given protograph
is lifted with an even lifting factor $\LF$ and coded bits are mapped
consecutively to a $4$-ary modulation. Then, $\LF/2$ VNs in each
class are allocated to the first modulation bit and $\LF/2$ to
the second. 

We model the bit mapper by specifying the assignment of \gls{vn}
classes to the bit channels via a matrix $\mathbf{A} = [a_{i,j}] \in
\mathbb{R}^{m \times n'}$, where $a_{i,j}$, $0 \leq a_{i,j} \leq 1$
$\forall i, j$ denotes the proportional allocation of \VNs from the
$j$th class (corresponding to the $j$th column in the base matrix)
allocated to the $i$th bit in the signal constellation.  The
approaches in \cite{Divsalar2005a, Jin2010, Nosratinia2011a} can be
recovered by considering only nonfractional assignments, \IE $a_{i,j}
\in \{0, 1\}$. In that case, \VNs of the original protograph
\cite{Divsalar2005a, Jin2010} or an intermediate protograph
\cite{Nosratinia2011a} are directly assigned to the modulation bits. 


We point out that, instead of interpreting $a_{i,j}$ as a
deterministic fraction of \VNs in a particular class allocated to a
particular channel, one should interpret $a_{i,j}$ as a probability,
and study the bit mapper as a probabilistic mapping device that
assigns coded bits randomly to channels, similar to \cite{Liu2006a}.
Under this assumption, one may argue that the \VNs belonging to a
certain class ``see'' an equivalent bit channel which is the average
of the original bit channels $f_{L_{i,k}|B_{i,k}}(l|b)$, weighted
according to the probabilities $a_{i,j}$. The MI of each equivalent
bit channel is a weighted average of the original channels' MI as
shown in the following lemma. 

\begin{lemma}
	Let $\{ f_{L_i|B_i}(l|b) : 1\leq i \leq m \}$ be a collection of
	symmetric LLR channels. Consider a new channel $f_{L|B}(l|b)$, where
	transmission takes place over the $i$th channel in the collection
	with probability $\alpha_i$ and $\sum_i \alpha_i = 1$. Then $I(L;B)
	= \sum_i \alpha_i I(L_i;B_i)$ for uniform input bits. 
\end{lemma}
\begin{MyProof}
	The channel $f_{L|B}(l|b)$ is a symmetric LLR channel.  The claim
	then follows from $f_{L|B}(l|b) = \sum_i \alpha_i
	f_{L_i|B_i}(l|b)$ and the fact that the \gls{mi} between the output
	of a symmetric LLR channel $f_{L|B}(l|b)$ and uniform input bits is
	$I(L; B) = 1 - \int_{-\infty}^{\infty} f_{L|B}(l|0) \log_2(1+e^{-l}) \,
	\D l$. 
\end{MyProof}

If we collect the \gls{mi} corresponding to the original $m$ symmetric
LLR channels in a vector $\vect{I}(\rho) = (I_1(\rho), \dots,
I_m(\rho))$, then, multiplying $\vect{I}(\rho)$ by $\vect{A}$ leads to
a vector $(\tilde{I}_1, \tilde{I}_2, \dots, \tilde{I}_{n'})$ with the \glspl{mi} corresponding
to the averaged bit channels as seen by the $n'$ \gls{vn} classes.
These averaged bit channels are modeled as symmetric Gaussian LLR
channels with parameters $(\sigma_1^2, \dots, \sigma_{n'}^2)$. In
particular, the \gls{pexit} analysis for nonbinary modulation is
obtained by changing the initialization step in line 3 of Algorithm 1
and assigning $\sigma_i^2 = J^{-1}(\tilde{I}_i)^2$, where the algorithm takes
$\vect{A}$ as an additional input to compute $\tilde{I}_i$ as described. 

In order to have a valid probabilistic assignment, all columns in
$\vect{A}$ have to sum to one and all rows in $\vect{A}$ have to sum
to $n'/m$, \IE we have $m n'$ equality constraints in total. The first
condition ensures that, asymptotically, all \VNs are assigned to a
channel, while the second condition ensures that all parallel channels
are used equally often. The set of valid assignment matrices is
denoted by $\mathcal{A}^{m \times n'} \subset \mathbb{R}^{m \times
n'}$.  In the case of punctured \VNs, the corresponding columns in
$\vect{A}$ are removed and $n'$ is interpreted as the number of
\emph{unpunctured} \VNs. 

\subsection{Optimization}

For a given bit mapper, \IE for a given assignment matrix
$\mathbf{A}$, an approximate decoding threshold $\rho^*(\mathbf{A})$
can be found using Algorithm 1 as follows. Fix a certain precision
$\precision$, target bit error probability $\pe$, and maximum number of
iterations $\maxIter$. Starting from some SNR $\rho$ where Algorithm 1
converges to a successful decoding, $S = 1$, iteratively decrease
$\rho$ by $\precision$ until the decoding fails. The smallest $\rho$
for which $S=1$ is declared as the decoding threshold
$\rho^*(\mathbf{A})$. For any $\rho \geq \rho^*(\mathbf{A})$, we
denote the number of iterations until successful decoding by
$\lsuccess$.

We are interested in optimizing $\mathbf{A}$ in terms of the decoding
threshold for a given protograph and modulation format. The
optimization problem is thus
\begin{align}
	\mathbf{A}_\text{opt} = \underset{\mathbf{A} \in \mathcal{A}^{m
	\times n'}}{\text{argmin}} \quad & \rho^*(\mathbf{A}),
	\label{eq:objective}
\end{align} 
where the baseline system realizes a mapping of coded bits to
modulation bits such that $a_{i,j} = 1/m$, $\forall i,j$,
\RevA{resulting in identical variances $\sigma_i^2$ for the equivalent
bit channels of all VN classes}. The
corresponding assignment matrix is denoted by $\Aunif$.  The search
space $\mathcal{A}^{m \times n'}$ can be regarded as a convex polytope
$\mathcal{P}$ in $p = (m-1)(n'-1)$ dimensions by removing the last row
and column in $\vect{A}$, replacing the equality constraints with
inequality constraints, and writing the matrix elements in a vector
$\vect{x} \in \mathbb{R}^p$ according to the prescription
$x_{(i-1)n'+j} = a_{i,j}$ for $1 \leq i \leq m-1$ and $1 \leq j \leq
n'-1$.  While the search space is convex, one can show by simple
examples that the objective function is nonconvex in $\mathcal{P}$.
In the following, we discuss ways to obtain good bit mappers with
reasonable effort.  We also remark that some of the optimization
approaches proposed previously in the context of bit mapper
optimization for irregular \gls{ldpc} codes are not necessarily
appropriate in our case due to the higher number of VN classes, \IE
they can be too complex (for example the iterative grid search in
\cite{Cheng2012}) or do not explore the search space efficiently
(simple hill climbing approaches as in \cite{Richter2007}).

First, as an alternative to directly optimizing the decoding
threshold, we iteratively optimize the convergence behavior in terms
of the number of iterations until successful decoding as follows.
Initialize $\rho$ to the decoding threshold for the baseline bit
mapper, \IE $\rho = \rho^*(\Aunif)$. Find $\Aopt$ such that it
minimizes the number of decoding iterations until convergence for the
given $\rho$, \IE
\begin{align}
	\Aopt = \underset{\mathbf{A} \in \mathcal{A}^{m \times n'}}{\text{argmin}} \quad
	\lsuccess. \label{eq:objective2}
\end{align}
For the found optimized $\Aopt$, calculate the new decoding threshold
$\rho^*(\Aopt)$. If the threshold did not improve, stop. Otherwise,
set $\rho = \rho^*(\Aopt)$ and repeat the optimization.  \RevA{The
above iterative approach was already used by the authors to find good
bit mappers for \gls{scldpc} codes in \cite{Hager2014} for parallel
\glspl{bec}. This approach is largely based on the ideas presented in
\cite[Sec.~IV]{Richardson2001a}, where optimized degree distributions
for irregular \gls{ldpc} codes are found.} The computational
complexity can be significantly reduced compared to the threshold
minimization \eqref{eq:objective}. However, it is not guaranteed to
be equivalent to a true threshold optimization, \IE in general
$\mathbf{A}_{\text{opt}} \neq \Aopt$. \RevC{We employ differential
evolution \cite{Storn1997} to solve the optimization problem in
\eqref{eq:objective2},} \RevA{which has been previously applied by
many authors in the context of irregular LDPC codes
\cite[p.~396]{Ryan2009}.} \RevC{Differential evolution is a solver for
unconstrained optimization problems and we briefly indicate how the
algorithm is modified to account for the constrained search space.
First, since $\mathcal{A}^{m \times n'}$ can be regarded as a convex
polytope, it is straightforward to take uniformly distributed points
for the initial population via standard random walk procedures
\cite{Kaufman1998}. Second, if the algorithm generates a trial point
$\mathbf{x}_t$ that lies outside the polytope, we apply the following
randomized bounce-back strategy. Let $\mathcal{L}$ be the line segment connecting
$\mathbf{x}_t$ and a random point inside the polytope, and let
$\mathbf{x}_i$ be the intersecting point of $\mathcal{L}$ and the
boundary of $\mathcal{P}$. We replace $\mathbf{x}_t$ with a point
taken randomly from $\mathcal{L}$, such that it lies in $\mathcal{P}$
and has at most a distance $d$ from $\mathbf{x}_i$, where $d$ is the
distance between $\mathbf{x}_i$ and $\mathbf{x}_t$. For a detailed
description of the algorithm itself and some guidelines regarding the
optimization parameter choice, we refer the reader to
\cite{Storn1997}. }

The optimization complexity is further reduced by constraining the
maximum number of iterations $\maxIter$. Practical systems commonly
operate with a relatively small number of \gls{bp} iterations. For
example, in Sec.~\ref{sec:Results}, we assume 50 \gls{bp} iterations, and
hence the decoding thresholds are optimized for the same number of
iterations. In the simulative verification, we have observed that the
performance of the finite-length codes assuming $50$ BP iterations is
generally better using a bit mapper that is also optimized for
$\maxIter = 50$ compared to, say, $\maxIter = 1000$, although the
differences were small.

Additionally, for \gls{scldpc} codes, we take advantage of the
structure of the optimized bit mappers for parallel BECs
\cite{Hager2014}, which show a certain form of periodicity. The
optimization complexity can then be reduced by assuming that the
optimal solution lies in a lower-dimensional subspace of
$\mathcal{P}$, defined by assignment matrices that take on a periodic
form as $\vect{A} = (\vect{A}', \vect{A}'', \vect{A}'', \cdots,
\vect{A}'', \vect{A}''')$, with $m \times V$ matrices $\vect{A}'$,
$\vect{A}''$, and $\vect{A}'''$, where $V$ is the periodicity factor.
If $V$ is chosen small enough, the dimensionality of the search space
(\IE $(m-1)(3V-1)$) can be substantially reduced, which generally
improves the convergence speed of the differential evolution
algorithm. 

\RevA{The methods and complexity reduction techniques described above
have been selected to obtain a good trade-off between final
performance and design complexity. In certain cases, for example the
considered AR4JA code in the next section, it could be possible to
further improve the performance at the expense of a higher design
complexity by directly targeting the decoding
threshold optimization \eqref{eq:objective} without the need for the
iterative optimization (albeit we expect the improvements to be
incremental). On the other hand, for the considered SC-LDPC code, the
iterative optimization and periodicity assumptions were critical to
maintain a reasonable design complexity, which is mainly due to the
very large number of protograph VNs. }

\section{Results and discussion}
\label{sec:Results}

In this section, we present and discuss numerical results, and
illustrate the performance gains that can be achieved by employing
optimized bit mappers. For the baseline systems, we use a consecutive
mapping of coded bits to modulation bits. Alternatively, one may use a
uniformly random mapping, which has the same expected performance.

In order to show the flexibility of the technique, we consider four
different scenarios, combining both modulation formats with one code
based on the AR4JA protographs and one \gls{scldpc} code, where the
lifting factor is $\LF = 3000$ in all cases. For simplicity, the codes
are randomly generated without further consideration of the graph
structure. The protograph lifting procedure can in principle be
combined with standard techniques to avoid short graph cycles that may
potentially lead to high error floors \cite[Ch.~6.3]{Ryan2009}.
Alternatively, an additional outer algebraic code may be assumed,
which removes remaining errors to achieve a required target BER of
$10^{-15}$. A rate $R=2/3$ code based on the AR4JA protograph for
$\ell = 1$ is used, which is denoted by $\Carja{}$. For the spatially
coupled case with $\TL = 30$, a code based on the protograph described
in Sec.~\ref{sec:scldpc} is used, which is denoted by $\Csc{}$. For
the given value of $\TL$, the design rate is $R(30) = 0.656$.  For the
AR4JA code, standard \gls{bp} decoding is assumed with $\maxIter =
50$, while for the \gls{scldpc} codes, we employ a \gls{wd} with $\WS
= 5$ and $\maxIter = 10$, which again amounts to a total of 50
iterations per decoded bit. We also tried other combinations of $\WS$
and $\maxIter$ with a similar total number of iterations and this
combination gave the best performance. For the bit mapper
optimization and in particular the \gls{pexit} analysis, we use the
same values for $\maxIter$ and $\WS$, and additionally $\pe =
10^{-5}$. The finite-length bit mappers are obtained via the rounded
matrix $\LF \vect{A}^*$ from which the index assignment of coded bits
to modulation bits is determined. 

\RevA{Notice that in all four scenarios, the approaches in
\cite{Divsalar2005a,Jin2010, Nosratinia2011a} are either not possible
(due to a mismatch between the number of protograph VNs and the number of
modulation bits) or not feasible (due to the large complexity of the
resulting optimization). As an example, the protograph corresponding
to $\Csc{}$ has 90 VNs and can be directly connected to the three
distinct bit channels of PM-$64$-QAM. This leads, however, to a very
large number of possible (nonfractional) connections between
protograph VNs and modulation bits. }

\subsection{Linear transmission}

\begin{figure}[t]
	\centering
	\subfloat[$\Carja{}$ with BP decoding and $\maxIter = 50$]{\includegraphics{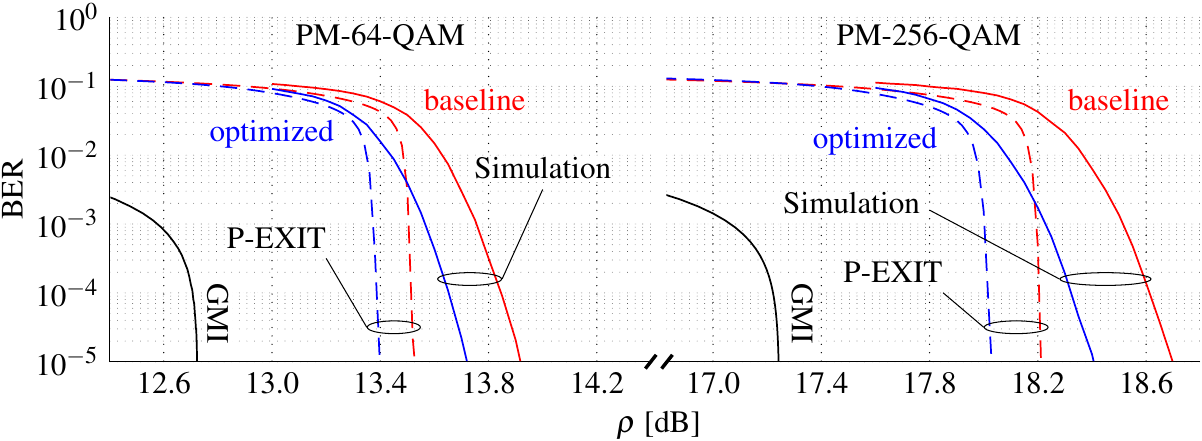}}

	\subfloat[$\Csc{}$ with a WD, $\WS = 5$, and $\maxIter = 10$]{\includegraphics{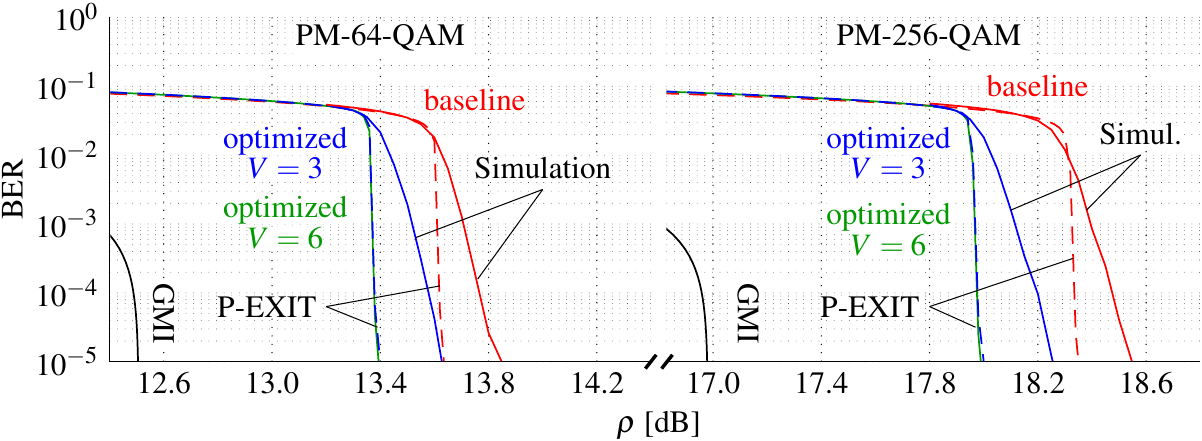}}

	\caption{Comparison of the optimized bit mappers (blue) with the
	baseline bit mappers (red) for the linear transmission
	scenario. Dashed lines correspond to P-EXIT analysis and solid lines
	to simulation results. In (b), solid green lines correspond to the P-EXIT
	analysis for $V=6$.}
	\label{fig:results_awgn}
\end{figure}

We start by providing a verification of the proposed optimization
technique assuming an \gls{awgn} channel. This case is obtained when
nonlinear effects are ignored, \IE $\gamma = 0$. In this case, the
channel \gls{pdf} \eqref{eq:gauss} is valid without approximations. 

In Fig.~\ref{fig:results_awgn}(a), the predicted \gls{ber} of the
AR4JA code
via the \gls{pexit} analysis is shown together with Monte Carlo
simulations by the dashed and solid lines, respectively.  Performance
curves for the baseline bit mappers are shown in red and for the
optimized ones in blue. As a reference, we also plot the
\gls{ber}-constrained \cite[p.~17]{Ryan2009} \gls{gmi} for the
corresponding spectral efficiency in each figure (the \gls{gmi} is
also referred to as the \gls{bicm} capacity \cite{Caire1998}). For
both scenarios, it can be observed that the optimized bit mappers lead
to a significant performance improvement. The gains that can be
achieved at a \gls{ber} of $10^{-5}$ are approximately $0.19$ and
$0.25$ dB for PM-64-QAM and PM-256-QAM, respectively. The predicted
gains from the P-EXIT analysis for the same \gls{ber} is slightly
less, \IE $0.12$ and $0.19$ dB, respectively. The deviation of the
asymptotic analysis from the actual simulation results is to be
expected due to the Gaussian approximation of the LLR densities and
the finite lifting factor and, hence, finite block lengths of the
codes. However, it is important to observe that, even though the
optimization was carried out assuming a cycle-free graph structure,
the predicted performance gains for the finite-length codes is well
preserved.

Similarly, the performance of the \gls{scldpc} code is shown in
Fig.~\ref{fig:results_awgn}(b). The periodicity factor for the bit
mapper optimization was set to $V = 3$. The observed gains at a
\gls{ber} of $10^{-5}$ are approximately $0.20$ dB for PM-64-QAM and
$0.25$ dB for PM-256-QAM. We also show the predicted P-EXIT
performance obtained for bit mappers that are optimized assuming a
larger periodicity factor of $V = 6$ by the solid green curves. It can be
seen that for both modulation formats, the additional gains are
incremental, \IE for PM-64-QAM the predicted performance curves
virtually overlap, while for PM-256-QAM, the difference is roughly
$0.01$ dB. This suggests that a full optimization of $\vect{A}$ will
be only marginally better than with $V = 3$.

\RevA{From Fig.~\ref{fig:results_awgn}, it appears that the P-EXIT
analysis consistently underestimates the finite-length performance
improvement for the AR4JA code, while it overestimates the improvement
for the \gls{scldpc} code. This observation does, however, not apply
in general and seems to be coincidental. In particular, we also
optimized the bit mapper for AR4JA codes of different code rates
(results not shown), and the P-EXIT analysis may also underestimate the true
performance improvements in that case.  Moreover,} we would like to
stress that a direct comparison between the two codes is difficult,
because of the slightly different code rates (and hence spectral
efficiencies) and different decoding complexities and delays.  Fair
comparisons between \gls{scldpc} codes and LDPC block codes is an
active area of research and beyond the scope of this paper. 
 
\subsection{Nonlinear transmission}

In this section, we consider a transmission scenario including
nonlinear effects, \IE $\gamma \neq 0$, where the assumed channel
\gls{pdf} \eqref{eq:gauss} is only approximately valid. In particular,
we study the potential increase in transmission reach that can be
obtained by employing the optimized bit mappers. 

We consider a single channel transmission scenario to keep the
simulations within an acceptable time. In the simulation model, we
assume perfect knowledge about the polarization state, and perfect
timing and carrier synchronization. All chosen system parameters are
summarized in Table~\ref{tab:parameters}.  Additionally, we use a
root-raised cosine pulse $p(t)$ with a roll-off factor of $0.25$. In
order to solve \eqref{eq:manakov}, we employ the symmetric \gls{ssfm}
with two samples per symbol and a fixed step size of $\Delta =
(10^{-4} L_{\text{D}}^2 L_{\text{NL}})^{1/3}$, where $L_{\text{D}} =
1/(|\beta_2| R_s^2)$ and $L_{\text{NL}} = 1/(\gamma P)$ is the
dispersive and nonlinear length, respectively. The input power that
maximizes $\rho$ according to the \gls{gn} model varies between $-2.2$
dBm for $\Nsp = 10$ and $-2.6$ dBm for $\Nsp = 40$. For simplicity,
the input power per polarization is fixed to $P = -2.5$ dBm for all
values of $\Nsp$. 

\begin{table}[t]
	\centering
	{\small
	\begin{tabular}{| c | c | l |}
		\hline
		\bfseries parameter& \bfseries meaning & \bfseries value \\\hline \hline
		$R_s$ & symbol rate & 40 Gbaud \\
		$\Lsp$ & span length & 70 km \\
		$\alpha$ & attenuation coefficient (0.25 dB/km) & 0.0576 km${ }^{-1}$ \\
		$\beta_2$ & chromatic dispersion coefficient & -21.668 ps${ }^2$/km \\
		$\gamma$ & nonlinear Kerr parameter & 1.4 W${ }^{-1}$ km${ }^{-1}$ \\
		$\nu_s$ & carrier frequency ($1550$ nm) & $1.934 \times 10^{14}$ Hz \\
		$\SEF$ & spontaneous emission factor & 1.622 \\ \hline
	\end{tabular}
	}
	\caption{System parameters}
	\label{tab:parameters}
\end{table}

\begin{figure}[t]
	\centering
	\includegraphics{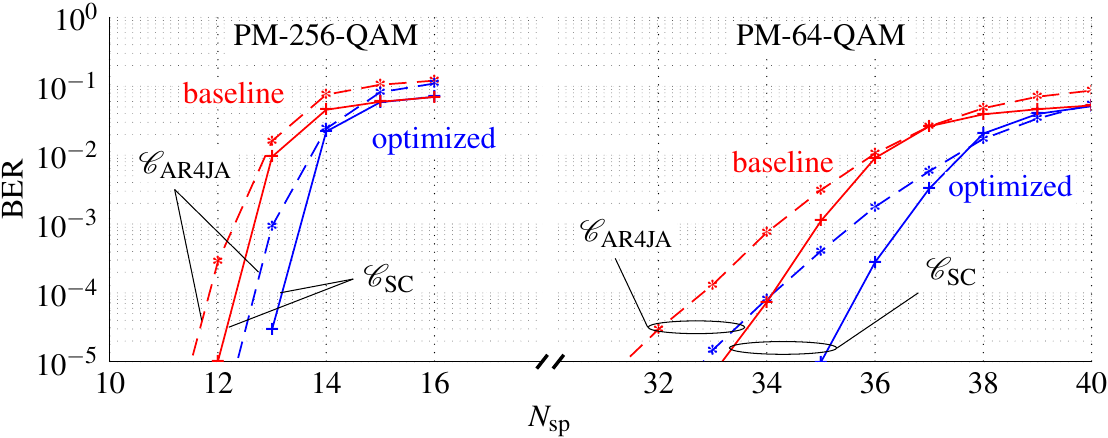}
	\caption{Comparison of the optimized bit mappers (blue) with the
	baseline bit mappers (red) for the nonlinear transmission
	scenario. }
	\label{fig:Reach}
\end{figure}

In Fig.~\ref{fig:Reach}, the simulated BER of the \gls{pm} systems
using $\Carja{ }$ and $\Csc{ }$ is shown as a function of the number
of fiber spans $\Nsp$ by the dashed and solid lines, respectively.
Again, curves corresponding to the baseline bit mappers are shown in
red, while curves corresponding to the optimized bit mappers are shown
in blue. Notice that the SNR decrease (in dB) is not linear with
increasing number of spans, hence the different slopes compared to the
curves shown in Fig.~\ref{fig:results_awgn}. For PM-256-QAM, the
transmission reach can be increased by roughly one additional span for
both codes, at the expense of a slightly increased BER. For example,
for $\Csc{ }$, the transmission reach can be increased from $12$ to
$13$ spans, while the BER slightly increases from $10^{-5}$ to $3
\cdot 10^{-5}$. For PM-64-QAM, the increase is roughly 1 span for
$\Carja{ }$ and roughly 2 spans for $\Csc{ }$. In fact, these gains
can be approximately predicted also from the GN model. For example,
for the chosen input power and system parameters, the GN model
predicts an SNR decrease of roughly 0.3 dB from $\Nsp=12$ to $\Nsp=13$
and 0.15 dB from $\Nsp=34$ to $\Nsp = 35$, \IE one would expect the
performance improvements in the linear transmission scenario to
translate into roughly one additional span for PM-256-QAM and one to
two additional spans for PM-64-QAM. This estimate corresponds to an
increase of the transmision reach by $3$--$8$\%, which is well in line with
the simulation results presented in Fig.~\ref{fig:Reach}.

\section{Conclusion}
\label{sec:Conclusion}

In this paper, we studied the bit mapper optimization for a \gls{pm}
fiber-optical system. Focusing on protograph-based codes, an
optimization approach was proposed based on a fractional allocation of
protograph bits to modulation bits via a modified \gls{pexit}
analysis. 
Extensive numerical simulations were used to verify the analysis for a
dispersion uncompensated link assuming both linear and nonlinear
transmission regimes. The results show performance improvements of up
to $0.25$ dB, translating into a possible extension of the
transmission reach by up to $8$\%.

\section*{Acknowledgments}

This work was partially funded by the Swedish Research Council under
grant \#2011-5961 and by the European Community's Seventh's Framework
Programme (FP7/2007-2013) under grant agreement No.~271986. The
simulations were performed in part on resources provided by the
Swedish National Infrastructure for Computing (SNIC) at C3SE.

\end{document}